\documentclass[letterpaper, 10pt, conference]{ieeeconf}

\IEEEoverridecommandlockouts                              

\usepackage{graphics} 
\usepackage{epsfig} 
\usepackage{mathptmx} 
\usepackage{times} 
\usepackage{amsmath} 
\usepackage{amssymb}  
\usepackage{bm}  
\usepackage{newtxtext, newtxmath}
\usepackage{color}
\usepackage{cite}
\usepackage{algorithm}
\usepackage{algpseudocode}
\usepackage{url}

\algnewcommand{\IfThen}[2]{
  \State \algorithmicif\ #1\ \algorithmicthen\ #2}

\algnewcommand{\IfThenElse}[3]{
  \State \algorithmicif\ #1\ \algorithmicthen\ #2\ \algorithmicelse\ #3}

\newcommand{\state}{z}
\newcommand{\dyn}{f}
\newcommand{\ctrl}{u}
\newcommand{\dstb}{d}

\newcommand{\cset}{\mathcal U}
\newcommand{\dset}{\mathcal D}
\newcommand{\cfset}{\mathbb U}
\newcommand{\dfset}{\mathbb D}

\newcommand{\raset}{\mathcal R}
\newcommand{\targetset}{\mathcal T}
\newcommand{\avoidset}{\mathcal A}
\newcommand{\roi}{\mathcal I} 

\newcommand{\astate}{x^\text A}
\newcommand{\dastate}{\dot x^\text A}
\newcommand{\dstate}{x^\text D}
\newcommand{\ddstate}{\dot x^\text D}
\newcommand{\apos}{p^\text A}
\newcommand{\dpos}{p^\text D}

\newcommand{\valfunc}{V}
\newcommand{\capradius}{R}

\newcommand{\lag}{L}
\newcommand{\issos}{\text{ is SOS}}
\newcommand{\setfunc}{\phi}
\newcommand{\fbctrl}{K}

\newtheorem{property}{Property}

\title{\Large \bf
Reach-Avoid Problems via Sum-of-Squares Optimization and \\Dynamic Programming
}

\author{Benoit Landry*, Mo Chen*, Scott Hemley, Marco Pavone	
\thanks{* These authors contributed equally to this paper.}
\thanks{B. Landry and M. Pavone are with the Department of Aeronautics and Astronautics, Stanford University. \{blandry, pavone\}@stanford.edu}
\thanks{M. Chen is with the School of Computing Science, Simon Fraser University. mochen@cs.sfu.ca}
\thanks{S. Hemley is with the Department of Mechanical Engineering, Stanford University. shemley@stanford.edu}%
}

\newcommand{\R}{\mathbb{R}}

\begin{document}

\maketitle
\thispagestyle{empty}
\pagestyle{empty}

\begin{abstract}
  Reach-avoid problems involve driving a system to a set of desirable configurations while keeping it away from undesirable ones. 
  Providing mathematical guarantees for such scenarios is challenging but have numerous potential practical applications.
  Due to the challenges, analysis of reach-avoid problems involves making trade-offs between generality of system dynamics, generality of problem setups, optimality of solutions, and computational complexity.
  In this paper, we combine sum-of-squares optimization and dynamic programming to address the reach-avoid problem, and provide a conservative solution that maintains reaching and avoidance guarantees.
  Our method is applicable to polynomial system dynamics and to general problem setups, and is more computationally scalable than previous related methods.
  Through a numerical example involving two single integrators, we validate our proposed theory and compare our method to Hamilton-Jacobi reachability.
  Having validated our theory, we demonstrate the computational scalability of our method by computing the reach-avoid set of a system involving two kinematic cars.
\end{abstract}

\section{Introduction}

Reach-avoid problems are prevalent in many engineering applications, especially those involving strategic or safety-critical systems. 
In these situations, one aims to find a control strategy that guarantees reaching a desired set of states while satisfying certain state constraints, all while accounting for unknown disturbances, which may be used to model adversarial agents. 
Reach-avoid sets capture the set of states from which the above task is \textit{guaranteed} to be successful. 
Reach-avoid problems are challenging to analyze due to the asymmetric goals of the control and disturbance, leading to non-convex, max-min cost functions \cite{Margellos11,Fisac15,Yang2017}. 
Due to the complexity of the cost function, dynamic programming-based methods for computing reach-avoid sets on a grid representing a state space discretization, such as Hamilton-Jacobi (HJ) formulations, have been popular and successful \cite{Mitchell05, Margellos11, Fisac15}. 

One specific class of reach-avoid problems is the reach-avoid game, in which the system consists of two adversarial players or teams. 
The first player, the attacker, assumes the role of the controller, and aims to reach some goal. 
The other player, the defender, assumes the role of the disturbance, and tries to prevent the attacker from achieving its goal. 
In \cite{Huang11, Huang2015}, the authors analyzed the two-player game of capture-the-flag by formulating it as a reach-avoid game, and then obtaining optimal strategies and winning regions for each player. 
The authors in \cite{Chen14, Chen17} extended previous results by analyzing the multiplayer case in which each team has an arbitrary number of players. 
Other dynamic programming-based methods for stochastic systems also exist \cite{MohajerinEsfahani2016, HomChaudhuri2016}. 
Other important applications of reach-avoid problems include motion planning in the presence of moving obstacles \cite{Vandenberg08,Wu2012,Giese2014,Chen2016d}. In particular, the multi-vehicle motion planning problem is analyzed from the perspective of reach-avoid problems and solved using dynamic programming in \cite{Chen2016d}.

\begin{figure}
  \centering
  \includegraphics[width=\columnwidth]{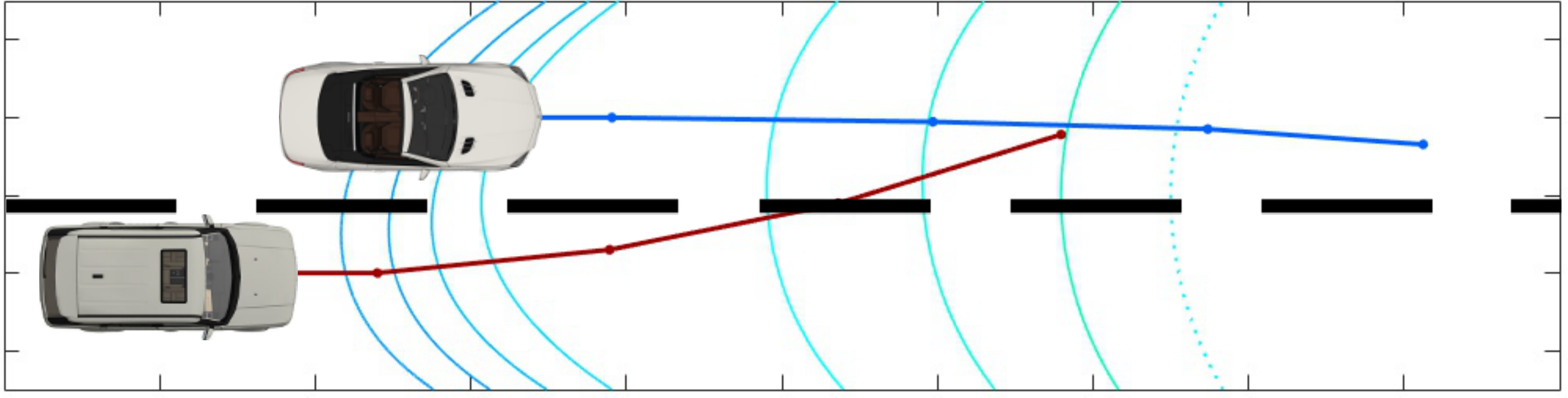}
  \caption{Example application of a reach-avoid closed-loop controller computed with our proposed algorithm. The controller of the top car guarantees it will reach the target region (dotted lines) without ever colliding with the bottom car, regardless of the bottom car's policy. In this case the bottom car is drifting into the top car's lane.}
  \label{fig:car_main}
\end{figure}

Due to the difficulties of analyzing reach-avoid problems, dynamic programming methods have enjoyed great success due to their optimality and generality when the state space is smaller than 6D.
For larger state spaces, computational burden becomes the main challenge.
To address this challenge, heuristics typically based on information pattern simplifications and multi-agent structural assumptions have been proposed. In \cite{Zhou2012}, the authors consider an open-loop formulation of the reach-avoid game in which one of the players declares its strategy at the beginning of the game; in addition, the player's dynamics are assumed to be single integrators in 2D. 
An open-loop framework is also used in \cite{Shih-YuanLiu2013} for pursuit-evasion. 
A semi-open-loop approach for 2D single integrator players, based on the idea of ``path defense,'' has been proposed in \cite{Chen2014}. 
In the context of multi-vehicle motion planning, priority-based heuristics have been used \cite{Chen2016d}. 
More broadly, methods for analyzing reach-avoid problems as well as related multi-agent systems make trade offs among generality of system dynamics, generality of the problem set up, conservatism of solutions, and computational efficiency \cite{Zhou2016, Kothari2017, Pierson2017, Xue2017, Zha2017, Yan2017, Li2016, Li2017}. 

Since the action of an opposing agent is modeled as a disturbance in the joint system, reach-avoid problems are closely related to robust planning, in which disturbance rejection is of primary concern. 
In this context, methods that produce value functions with Lyapunov-like properties have been very effective.
When the system dynamics are nonlinear in general but have small state spaces, HJ methods are able to produce Lyapunov-like functions through dynamic programming \cite{Mitchell05, Fisac15}. 
Computational complexity has been alleviated in specific scenarios using decomposition techniques \cite{Chen2016b,Chen2016a}; however, these are not applicable to general reach-avoid problems.

When the system dynamics and functions representing sets are polynomial, the search for Lyapunov-like functions can be done more efficiently by leveraging sum-of-squares (SOS) programs, which can be converted to semidefinite programs \cite{Parrilo2000} and solved using standard optimization toolboxes \cite{mosek2010mosek}.
SOS programs involve checking whether polynomial functions can be written as SOS, which is a sufficient condition for non-negativity or positivity, and thereby establishing Lyapunov-like properties.
In addition, complex problem statements involving sets and implications can be written as SOS constraints.
SOS programs have been used extensively in robust planning with methods involving barrier certificates and robust funnels \cite{Tedrake2010, Barry2012, Majumdar2013, Majumdar2017}.
Other methods that utilize nonlinear optimization also exist; for example, \cite{Gao2017} and \cite{Alonso-Mora2017} utilize nonlinear optimization techniques for motion planning through dynamic environments for a single vehicle and a flock of vehicles, respectively.
  
\textit{Statement of contributions:} 
In this paper, we propose a method for computing the reach-avoid set and synthesizing a feedback controller that is \textit{guaranteed} to drive the system into a target set while staying out of an avoid set.
Our approach combines dynamic programming and SOS optimization: the reach-avoid set, represented by a Lyapunov-like value function, is obtained backwards one time step at a time as in dynamic programming, and at each time step a SOS program is solved.
Building on previous SOS-based work such as \cite{Majumdar2013,Majumdar2017}, we explicitly encode the avoidance constraint so that a single value function guarantees both reaching and avoidance, as in HJ methods \cite{Zhou2012, Huang2015, Chen17}.
Compared to previous dynamic programming-based work such as \cite{Zhou2012, Huang2015, Chen17}, we trade off optimality of solution for computational complexity: although our method produces conservative reach-avoid sets, we are able to analyze systems with higher-dimensional state spaces.
Unlike analytic approaches such as \cite{Chen2014, Pierson2017, Zha2017}, our approach applies much more broadly to different problem setups and system dynamics.
We demonstrate our method in simulations.
  
  The remainder of this paper is structured as follows: 
  \begin{itemize}
    \item In Section \ref{sec:bg}, we formulate the reach-avoid problem and provide some background about SOS programming.
    \item In Section \ref{sec:sos}, we derive the SOS constraints for computing the reach-avoid set based on its basic properties.
    \item In Section \ref{sec:dp}, we propose a dynamic programming approach for solving the SOS program more efficiently.
    \item In Section \ref{sec:num}, we numerically validate our theory by comparing our method to the HJ method, and present, to the best of our knowledge, the first 6D reach-avoid set computed using a general numerical method.
    \item In Section \ref{sec:conclusion}, we conclude and provide suggestions for future directions.
  \end{itemize}
\section{Preliminaries}\label{sec:bg}
\subsection{Problem Formulation}
Consider a system which evolves according to its dynamics, given by the following ordinary differential equation:

\begin{equation} \label{eq:dyn}
\dot\state=\dyn(\state,\ctrl,\dstb),\qquad \ctrl \in \cset, \dstb \in \dset
\end{equation}

\noindent where $\state \in \R^n$ is the system state, $\ctrl \in \cset$ is its control, and $\dstb \in \dset$ is the disturbance. We assume that the control  must be of a time-varying state feedback form, $\ctrl = \fbctrl(s, \state)$, and that $\ctrl(\cdot)$ and $\dstb(\cdot)$ are measurable. Importantly, we make no assumption on the disturbance $\dstb$ other than that it is bounded. We denote the function spaces from which $\ctrl(\cdot)$ and $\dstb(\cdot)$ are drawn as $\cfset$ and $\dfset$, respectively.

The system dynamics $\dyn: \R^n \times \cset \times \dset \rightarrow \R^n$ is assumed to be uniformly continuous, bounded, and Lipschitz continuous in $\state$ for fixed $\ctrl$ and $\dstb$. So given $\ctrl(\cdot) \in \cfset, \dstb(\cdot) \in \dfset$, there exists a unique trajectory solving \eqref{eq:dyn} \cite{Coddingtona}. 


We would like to compute the set of joint states from which the attacker wins the game of time horizon $|t|,t\le0$. This is captured by the reach-avoid set, defined as follows:

\begin{align}
&\raset(t) = \{\state: \exists \ctrl(\cdot) \in \cfset, \forall \dstb(\cdot) \in \dfset, \text{ such that } \label{eq:ra_def}\\
&\qquad \state(\cdot) \text{ satisfies } \eqref{eq:dyn}, \exists s \in [t, 0], \state(s) \in \targetset \wedge \nonumber\\ 
&\qquad \forall \tau \in [t, s), \state(\tau) \notin \avoidset\} \nonumber
\end{align}

The avoid set $\avoidset$ is the set of states that the system must avoid while reaching the target. Note the following propety:

\begin{property} \label{prop:fc}
  \textbf{Final condition.} $\raset(0) = \targetset$.
\end{property}

\subsection{SOS Programming Background} \label{subsec:sos}
In this section, we provide a brief introduction to SOS programs. For a more detailed discussion, please refer to \cite{Parrilo2000} and \cite{Majumdar2017}. 
%
In this paper we will represent a set $\mathcal G$ of states $x = (x_1, \ldots, x_n) \in \R^n$ as $\phi_\mathcal G(x)$: $\mathcal G = \{x : \phi_\mathcal G(x) \le 0\}$. This allows us to transform set-based constraints such as $x \in \mathcal G$ to constraints of the form $\phi_\mathcal G(x) \le 0$ in our proposed optimization problem. Note that such a constraint is generally non-convex. 

When $\phi_\mathcal G(x)$ is a polynomial in $x$, checking non-positivity over $\R^n$ is still NP-hard \cite{Parrilo2000}. However, in this case the constraint $\phi_\mathcal G(x) \le 0$ can be relaxed to the SOS condition $-\phi_\mathcal G(x) = \sum_{i=1}^l p_i^2 (x)$, where $p_i(x)$ are polynomials. This condition is equivalent to $-\phi_\mathcal G(x) = \nu^\top(x) Q \nu(x), Q \succeq 0$, where $\nu(x)$ is a vector of polynomial basis functions up to less than or equal to half the degree of $\phi_\mathcal G(x)$, and $Q$ is a semi-definite matrix of appropriate size. Note that this constraint is satisfied by matching coefficients on the left- and right-hand sides. A short-hand for the above constraint is ``$-\phi_\mathcal G(x) \text{ is SOS}$''.

One is often interested in guaranteeing non-positivity over a subset of state space. For example, one may desire $\phi_A(x)\le 0$ over the set where $\phi_B(x) \le 0$. A constraint in the form of $x \in B \Rightarrow \phi_A(x) \le 0$  where $B = \{x: \phi_B(x) \le 0\}$ can be written as $-\phi_A(x) + L(x)\phi_B(x) \issos, L(x) \issos$. 

%
\section{Solution via SOS Programming}\label{sec:sos}
We now formulate a SOS program whose solution characterizes the reach-avoid set defined in Eq. \eqref{eq:ra_def}, and provides a feedback controller that guarantees reaching and avoidance. 

As in earlier literature \cite{Majumdar2017}, let $\raset(s)$ be characterized by the $\rho(s)$ sublevel set of some function $\valfunc(s, \state)$:

\begin{equation} \label{eq:sublevel}
\raset(s) = \{\state: \valfunc(s, \state) \le \rho(s)\}
\end{equation}

%
%
%
%

\subsection{The Value Function}
Our SOS formulation is motivated by basic properties of the reach-avoid set. The first is Property \ref{prop:fc} in Section \ref{sec:bg}. Taking the convention that $\state \in \targetset \Leftrightarrow \setfunc_\targetset(\state) \le 0$, this property can be stated as $\setfunc_\targetset(\state) = \valfunc(0, \state) - \rho(0)$.

%
%

\noindent
\begin{property}\label{prop:vdot} 
\textbf{Lyapunov-like property.} By definition of $\raset(s)$, if $\state(s) \in \raset(s)$ and $\state(s) \notin \targetset$, then there exists $\ctrl(s)$ such that for all $\dstb(s)$, $\state(s+\epsilon) \in \raset(s+\epsilon)$ for any arbitrarily small $\epsilon>0$. 
\end{property}

In terms of the value function, and with the convention $\dstb\in\dset \Leftrightarrow\setfunc_\dset(\dstb) < 0$, Property \ref{prop:vdot} becomes

\begin{align*}
&\forall s \in [t, 0], \state \in \R^n,  \\
&\setfunc_\targetset(\state) > 0 \wedge \setfunc_\dset(\dstb) \le 0 \wedge \valfunc(s,\state) = \rho(s)\Rightarrow \dot \valfunc(s,\state) \le \dot\rho(s)
\end{align*}


\noindent where 
\begin{equation} \label{eq:vdot}
\dot \valfunc(s,\state) = \frac{\partial\valfunc(s,\state)}{\partial\state} \cdot \dyn(\state, \ctrl, \dstb) + \frac{\partial \valfunc(s,\state)}{\partial s}
\end{equation}


This Lyapunov-like property states that if $\state(s)$ is not in $\targetset$ and $\state(s)$ is on the boundary of $\raset(s)$, then there must some control $\ctrl$, over which the SOS program will optimize, such that regardless of the chosen disturbance $\dstb\in\dset$, $\valfunc(s, \state(s))$ will remain non-positive. The boundary of $\raset(s)$ is described by the condition $\valfunc(s,\state) = 0$, and the non-positivity condition on $\dot\valfunc - \dot\rho$ ensures the continued non-positivity of $\valfunc-\rho$.

\begin{property} \label{prop:avoid}
  \textbf{Avoidance property.} By the definition of $\raset(t)$, if $\state(s)\in\avoidset$ and $\state(s)\notin\targetset$, then it cannot be in $\raset(s)$.
\end{property} 

Given the Eq. \ref{eq:sublevel}, Property \ref{prop:avoid} is equivalent to the following:

\begin{equation*}
\forall s \in [t, 0], \state \in \R^n, ~ \setfunc_\targetset(\state) > 0\wedge \setfunc_\avoidset(\state) \le 0 \Rightarrow \valfunc(s,\state) > \rho(s)
\end{equation*}

%

%


\subsection{Control Parametrization and Bounds}
Equation \eqref{eq:vdot} depends on the control $\ctrl$, which is bounded according to the system dynamics \eqref{eq:dyn}. By the definition of the reach-avoid set in Eq. \eqref{eq:ra_def}, we must take this into account. 

In this paper, we aim to search for a feedback controller $\ctrl = \fbctrl(s, \state)$, so following the control bounds constraint in \cite{Majumdar2013} and \cite{Majumdar2017}, we enforce the following constraint:

\begin{align}
\forall s \in [t,0], \state \in \R^n, ~\valfunc(s, \state) \le \rho(s) \wedge \setfunc_\dset(\dstb) \le 0 \nonumber\\
 \Rightarrow \setfunc_\cset(\fbctrl(s, \state)) \le 0
\end{align}

As long as $\setfunc_\cset(\cdot)$ is linear, the set $\{\state: \setfunc_\cset(\fbctrl(s, \state)) \le 0\}$ is semi-algebraic for a given $s$.


%
%
%
%
%


\subsection{The SOS Program}

Putting all of the above into consideration, we arrive at the following optimization problem in Eq. \eqref{eq:opt0:obj}, which maximizes the volume of the reach-avoid set while enforcing the constraints described above. We will describe how volume can be maximized in Section \ref{sec:dp:obj}.

  \begin{align}
  \underset{\valfunc(\cdot, \cdot),\fbctrl(\cdot,\cdot), \rho(\cdot)}{\text{maximize}}\quad &\int_{s=t}^0\text{volume}(\raset(s))ds \label{eq:opt0:obj} \\
  \text{subject to}\quad & \setfunc_\targetset(\state) = \valfunc(0, \state) - \rho(0) \nonumber
  \\
  &\forall s \in [t,0], \state \in \R^n, \nonumber\\
  &\quad 
  \begin{aligned}
  (\setfunc_\targetset(\state) > 0 \wedge \setfunc_\dset(\dstb) \le 0 \wedge \valfunc(s,\state) = \rho(s)) \\
  \Rightarrow \dot \valfunc(s,\state) \le \dot\rho(s)
  \end{aligned} \nonumber
  \\
  &\quad  (\setfunc_\targetset(\state) > 0\wedge \setfunc_\avoidset(\state) \le 0) \Rightarrow \valfunc(s,\state) > \rho(s) \nonumber 
  \\
  &\quad \begin{aligned}
   (\valfunc(s, \state) \le \rho(s) \wedge \setfunc_\dset(\dstb) \le 0) \\
   \Rightarrow \setfunc_\cset(\fbctrl(s, \state)) \le 0
  \end{aligned} \nonumber
\end{align}

As introduced in Section \ref{subsec:sos}, we can convert all the above constraints into SOS form. Plugging in Eq. \eqref{eq:vdot} for $\dot\valfunc$, $\ctrl = \fbctrl(s,\state)$, the optimization program becomes \eqref{eq:opt1:obj}:

  \begin{align}
  \underset{\valfunc,\fbctrl, \rho, \mathbf{\lag}}{\text{maximize}}\quad &\int_{s=t}^0\text{volume}(\raset(s))ds \label{eq:opt1:obj} \\
  \text{subject to}\quad & \setfunc_\targetset(\state) = \valfunc(0, \state) - \rho(0) \nonumber
  \\
  &\forall s \in [t,0], \state \in \R^n, \nonumber\\
  &\qquad 
  \begin{aligned}
   &-\Big(\frac{\partial\valfunc(s,\state)}{\partial\state} \cdot \dyn(\state, \fbctrl(s,\state), \dstb) \\
   & \qquad + \frac{\partial \valfunc(s,\state)}{\partial s} - \dot\rho(s)\Big) + \lag_\dset^\text{Lyap}(s,\state, \dstb) \setfunc_\dset(\dstb)\\
   & \qquad\qquad\qquad+ \lag_\raset^\text{Lyap}(s,\state,\dstb)(\valfunc(s,\state) - \rho(s))  \\
   & \qquad\qquad\qquad- \lag_\targetset^\text{Lyap}(s,\state,\dstb)\setfunc_\targetset(\state) \issos \\
  \end{aligned} \nonumber
  \\
  &\qquad 
  \begin{aligned}
  \valfunc(s,\state) - \rho(s) &- \lag_\targetset^\text{ca}(s,\state)\setfunc_\targetset(\state) \\
  &+ \lag_\avoidset^\text{ca}(s,\state)\setfunc_\avoidset(\state) \issos 
  \end{aligned}\nonumber
  \\
  &\qquad 
  \begin{aligned}
  -\setfunc_\cset(\fbctrl(s,\state)) &+ \lag_\raset^\text{ctrl}(s,\state,\dstb)(\valfunc(s, \state) - \rho(s)) \\
  &+ \lag_\dset^\text{ctrl}(s,\state,\dstb)\setfunc_\dset(\dstb) \issos
  \end{aligned} \nonumber
  \\
  &\begin{aligned}
  &\lag_\dset^\text{Lyap}, \lag_\targetset^\text{Lyap}, \lag_\targetset^\text{ca}, \lag_\avoidset^\text{ca},\lag_\raset^\text{ctrl}, \lag_\dset^\text{ctrl} \text{ are SOS}
  \end{aligned} \nonumber
  \end{align}

\noindent where $\mathbf L = \big\{\lag_\dset^\text{Lyap}, \lag_\targetset^\text{Lyap}, \lag_\raset^\text{Lyap}, \lag_\targetset^\text{ca}, \lag_\avoidset^\text{ca}, \lag_\raset^\text{ctrl}, \lag_\dset^\text{ctrl}\big\}$.

\section{Solving the SOS Program via Dynamic Programming}\label{sec:dp}

The optimization in \eqref{eq:opt1:obj} involves polynomials in continuous time and state space. In this section, we discretize time so that the problem can be solved in a dynamic programming fashion, one time step at a time, akin to what is done in HJ reachability. This way, we avoid optimizing over an entire time horizon, which is computationally expensive.

\subsection{Time Discretization}
We define time samples $\{s_k\}_{k = 0}^N$. All quantities dependent on time are now indexed by $k =\{0,\ldots,N\}$: for example, $\valfunc(s_k, \state) \approx \valfunc_k(\state)$ and $\raset(s_k) \approx \raset_k$. In addition, we approximate\footnote{Discretization error can be reduced with higher order schemes.} the derivatives of $\valfunc$ and $\rho$ with respect to $s$:

\begin{align}
\frac{\partial \valfunc(s_k, \state)}{ \partial s} \approx \frac{\valfunc_{k+1}(\state)-\valfunc_k(\state)}{s_{k+1} - s_k}, \quad \dot\rho(s) \approx \frac{\rho_{k+1}-\rho_{k}}{s_{k+1} - s_k}
\end{align}


Now, the optimization problem becomes \eqref{eq:opt2:obj}.

\begin{align}
\underset{\{\valfunc_k,\fbctrl_k, \rho_k, \mathbf{\lag}_k\}_{k=0}^{N-1}}{\text{maximize}}\quad &\sum_{k=0}^{N-1}\text{volume}(\raset_k) \label{eq:opt2:obj} 
\\
\text{subject to}\quad & \setfunc_\targetset(\state) = \valfunc_N(\state) - \rho_N \nonumber
\\
&\forall k \in 0, 1, \ldots, N-1, \state \in \R^n, \nonumber
\\
&\quad 
\begin{aligned}
&-\Big(\frac{\partial\valfunc(s,\state)}{\partial\state} \cdot \dyn(\state, \fbctrl_k(\state), \dstb) \\
& \quad + \frac{\valfunc_{k+1}(\state)-\valfunc_k(\state)}{s_{k+1} - s_k} - \frac{\rho_{k+1}-\rho_{k}}{s_{k+1} - s_k}\Big) \\
& \quad\qquad + \lag_{\dset,k}^\text{Lyap}(\state,\dstb) \setfunc_\dset(\dstb)\\
& \quad\qquad + \lag_{\raset,k}^\text{Lyap}(\state,\dstb)(\valfunc_k(\state) - \rho_k)  \\
& \quad\qquad - \lag_{\targetset,k}^\text{Lyap}(\state,\dstb)\setfunc_\targetset(\state) \issos \\
\end{aligned} \nonumber
\\
&\quad 
\begin{aligned}
\valfunc_k(\state) - \rho_k &- \lag_{\targetset,k}^\text{ca}(\state)\setfunc_\targetset(\state) \\
&+ \lag_{\avoidset,k}^\text{ca}(\state)\setfunc_\avoidset(\state) \issos 
\end{aligned}\nonumber
\\
&\quad 
\begin{aligned}
-\setfunc_\cset(\fbctrl_k(\state)) &+ \lag_{\raset,k}^\text{ctrl}(\state,\dstb)(\valfunc_k(\state) - \rho_k) \\
&+ \lag_{\dset,k}^\text{ctrl}(\state,\dstb)\setfunc_\dset(\dstb) \issos
\end{aligned}\nonumber 
\\
&\begin{aligned}
&\lag_{\dset,k}^\text{Lyap}, \lag_{\targetset,k}^\text{Lyap}, \lag_{\targetset,k}^\text{ca}, \lag_{\avoidset,k}^\text{ca}, \lag_{\raset,k}^\text{ctrl},\lag_{\dset,k}^\text{ctrl} \text{ are SOS}
\end{aligned}\nonumber 
\end{align}

\noindent where $\mathbf \lag_k = \big\{\lag_{\dset,k}^\text{Lyap}, \lag_{\targetset,k}^\text{Lyap}, \lag_{k,\raset}^\text{Lyap}, \lag_{\targetset,k}^\text{ca}, \lag_{\avoidset,k}^\text{ca}, \lag_{\raset,k}^\text{ctrl}, \lag_{\dset,k}^\text{ctrl}\big\}$.

\subsection{Dynamic Programming}
The optimization in \eqref{eq:opt2:obj} involves decision variables in the entire time horizon. However, the structure of the optimization program allows us to break it down into smaller problems, each representing one time step. This allows the computational complexity of our proposed method to scale linearly with the number of time discretization points.

Given $\setfunc_\targetset(\state) = \valfunc_N(\state) - \rho_N$ or $\valfunc_N(\state) = \setfunc_\targetset(\state) + \rho_N$ with some arbitrary $\rho_N$, the optimization program starts at $k=N-1$ and decrements $k$ after every optimization of the form


  \begin{align}
  \underset{\valfunc_k,\fbctrl_k, \rho_k, \mathbf{\lag}_k}{\text{maximize}}\quad &\text{volume}(\raset_k) \label{eq:optk:obj} \\
  \text{subject to}\quad & 
  \begin{aligned}
  &\forall \state \in \R^n, \nonumber\\
  &-\Big(\frac{\partial\valfunc(s,\state)}{\partial\state} \cdot \dyn(\state, \fbctrl_k(\state), \dstb) \\
  & \qquad + \frac{\valfunc_{k+1}(\state)-\valfunc_k(\state)}{s_{k+1} - s_k} - \frac{\rho_{k+1}-\rho_{k}}{s_{k+1} - s_k}\Big) \\
  & \qquad\qquad\qquad + \lag_{\dset,k}^\text{Lyap}(\state,\dstb) \setfunc_\dset(\dstb)\\
  & \qquad\qquad\qquad + \lag_{\raset,k}^\text{Lyap}(\state,\dstb)(\valfunc_k(\state) - \rho_k)  \\ 
  & \qquad\qquad\qquad - \lag_{\targetset,k}^\text{Lyap}(\state,\dstb)\setfunc_\targetset(\state) \issos \\
  \end{aligned} \nonumber
  \\
  & 
  \begin{aligned}
  \valfunc_k(\state) - \rho_k &- \lag_{\targetset,k}^\text{ca}(\state)\setfunc_\targetset(\state) \\
  &+ \lag_{\avoidset,k}^\text{ca}(\state)\setfunc_\avoidset(\state) \issos 
  \end{aligned}\nonumber
  \\
  & 
  \begin{aligned}
  -\setfunc_\cset(\fbctrl_k(\state)) &+ \lag_{\raset,k}^\text{ctrl}(\state,\dstb)(\valfunc_k(\state) - \rho_k) \\
  &+ \lag_{\dset,k}^\text{ctrl}(\state,\dstb)\setfunc_\dset(\dstb) \issos
  \end{aligned} \nonumber
  \\
  &\lag_{\dset,k}^\text{Lyap}, \lag_{\targetset,k}^\text{Lyap}, \lag_{\targetset,k}^\text{ca}, \lag_{\avoidset,k}^\text{ca}, \lag_{\raset,k}^\text{ctrl}, \lag_{\dset,k}^\text{ctrl} \text{ are SOS.}
  \nonumber
  \end{align}
  
  \noindent where $\mathbf \lag_k = \big\{\lag_{\dset,k}^\text{Lyap}, \lag_{\targetset,k}^\text{Lyap}, \lag_{k,\raset}^\text{Lyap}, \lag_{\targetset,k}^\text{ca}, \lag_{\avoidset,k}^\text{ca}, \lag_{\raset,k}^\text{ctrl}, \lag_{\dset,k}^\text{ctrl}\big\}$.


%

\subsection{Volume Maximization} \label{sec:dp:obj}
Maximizing the volume of a polynomial sublevel set is potentially intractable \cite{dabbene2012minimum}. We therefore substitute the objective of the previously defined optimization problems with a heuristic.
Using cost heuristics does not remove any guarantees from our approach.

Since $\valfunc_k(z) \le \rho_k$ represents $\raset_k$, one heuristic for maximizing volume is to restrict $\valfunc_k(z)$ to be SOS, minimize the integral of $\valfunc_k(\state)$ over a region of interest $\roi$ as in \cite{dabbene2012minimum}, and maximize $\rho_k$ as in \cite{Majumdar2013}. 
We will later also use $\roi$ to slightly relax the SOS program in Section \ref{sec:dp:imp}. 
Similarly to what is described in \cite{dabbene2012minimum}, we write $\valfunc_k(z) = c_k^V \cdot \nu(\state)$ where $c_k^V$ represents the coefficients of $\valfunc_k(z)$ and $\nu(\state)$ a monomial basis. 
This allows us to write the integral as a linear function of $c_k^V$, making it amenable to SOS optimization:

\begin{align} \begin{subequations}
\int_{\mathcal{I}}{\valfunc(s, \state)}d\state = c_k^V \cdot \int_{\mathcal{I}}{\nu(\state)}d\state
\end{subequations} \end{align}

If $\roi$ is a bounding box, the integral can be computed analytically. This volume maximization heuristic is reflected in \eqref{eq:optf:obj} and \eqref{eq:optf:SOS}. Note that the integral can only be minimized when $\valfunc_k$ is held fixed, otherwise a normalization constraint on the value function must be introduced \cite{Majumdar2013}.


\subsection{Implementation Details}\label{sec:dp:imp}

Since the optimization \eqref{eq:optk:obj} is bilinear and thus non-convex, we propose heuristics for obtaining useful solutions. This section describes the heuristics, provides the final optimization program \eqref{eq:optf}, and presents Alg. \ref{alg:opt} for solving it.

\noindent\textbf{Value function invariance}:
One property that also arises from the definition of $\raset(t)$ in Eq. \eqref{eq:ra_def} is as follows:

\begin{property}
  \textbf{Invariance property}. 
  \begin{equation*}
    \state \in \raset(t) \Rightarrow \forall t' < t, ~ \state \in \raset(t')
  \end{equation*}
\end{property}

In terms of the value function, this implies that $\valfunc_{k+1} \le \rho_{k+1} \Rightarrow \valfunc_{k} \le \rho_{k}$. Using a slack variable $\epsilon_k^\text{it} \ge 0$, we encoded this property as an additional soft constraint in Eq. \eqref{eq:optf:it} to guide the optimization. 
The value of $\epsilon_k^\text{it}$ is minimized in the objective \eqref{eq:optf:obj} with weight $\lambda_k^\text{it}$.
In our experience, without $\epsilon_k^\text{it}$, the optimization does not reliably produce value functions $\valfunc_k$ that represent non-empty sets.

\noindent\textbf{Region of interest}: 
To facilitate the search for polynomials $(\valfunc_k, \fbctrl_k)$ and constant $\rho_k$, we relax the constraints so that they only apply in some region of interest $\roi$, which can be chosen to either be large enough to contain the reach-avoid set $\{\raset_k\}$, or in general contain the region of the state space that one wishes to consider. This is done by adding the terms $\lag_{\roi,k}^\text{Lyap} \setfunc_\roi$, $\lag_{\roi,k}^\text{ca} \setfunc_\roi$, $\lag_{\roi,k}^\text{ctrl} \setfunc_\roi$, $\lag_{\roi,k}^\text{it} \setfunc_\roi $ in the constraints \eqref{eq:optf:lyap}, \eqref{eq:optf:ca}, \eqref{eq:optf:ctrl_sat}, \eqref{eq:optf:it}, respectively.

\noindent\textbf{Initialization}:
To obtain a feasible initial guess, we introduce a slack variable $\epsilon_k^\text{Lyap}$ to allow $\dot\valfunc$ to be initially non-negative in \eqref{eq:optf:lyap}. 
Throughout alternations of the optimization, we minimize $\epsilon_k^\text{Lyap}$ in the objective \eqref{eq:optf:obj}. 
In addition, we increase the weight $\lambda_k^\text{Lyap}$ by some factor $\alpha$ to drive the value of $\epsilon_k^\text{Lyap}$ down, as described in Alg. \ref{alg:opt}.
Once $\epsilon_k^\text{Lyap}$ is below some threshold $\delta_k^\text{slack}$, we consider the solution to be numerically feasible and stop optimizing over $\epsilon_k^\text{Lyap}$. 
In practice, we are consistently able to obtain feasible solutions.

\noindent\textbf{Final optimization problem}:
All of the above considerations lead to the final optimization program for each time step in \eqref{eq:optf}. The bilinear terms are $\valfunc_k \fbctrl_k$ and $\lag_{\raset,k}^\text{Lyap}(\valfunc_k - \rho_k)$ in \eqref{eq:optf:lyap},  $\lag_{\raset,k}^\text{ctrl}(\valfunc_k - \rho_k)$ in \eqref{eq:optf:ctrl_sat}, and $\lag_{\targetset,k}^\text{it}(\state) \epsilon_k^\text{it}$ in \eqref{eq:optf:it}. Therefore, we can optimize the three sets of variables $\{\epsilon_k^\text{Lyap}, \mathbf \lag_k\}$, $\{\valfunc_k, \mathbf \epsilon_k, \mathbf{\hat\lag}_k\}$, and $\{\fbctrl_k, \rho_k, \mathbf \epsilon_k, \mathbf{\hat \lag}_k\}$ in an alternating fashion, where $\mathbf{\hat\lag}_k:=\mathbf{\lag}_k\backslash\{\lag_{\raset,k}^\text{Lyap}, \lag_{\raset,k}^\text{ctrl}, \lag_{\avoidset,k}^\text{ca}, \lag_{\targetset,k}^\text{it}\}$. Note that as mentioned in Section \ref{sec:dp:obj}, $\rho_k$ cannot be optimized at the same time as $\valfunc_k$ due to the volume maximization heuristic. The optimization algorithm is outlined in Algorithm \ref{alg:opt}.

\begin{algorithm}[]
  \caption{}
  \label{alg:opt}
  \begin{algorithmic}[1]
    \State {\bf Input:} Initial guess $(\valfunc_k, \rho_k, \fbctrl_k, \epsilon_k^\text{it})$; slack weights $\mathbf \lambda_k$; weight growth rate $\alpha$; tolerances $\delta^\text{slack}, \delta^\text{conv}$; max. \# of iterations maxIter.
    \State $i \leftarrow 0$.
    \While {$i<$maxIter} 
    \State $(\valfunc_k^\text{prev}, \rho_k^\text{prev}, \fbctrl_k^\text{prev}) \leftarrow (\valfunc_k, \rho_k, \fbctrl_k)$
    \State Solve~\eqref{eq:optf} w.r.t. $\{\epsilon_k^\text{Lyap}, \mathbf \lag_k\}$ \label{subprob_1}
    \IfThenElse{$\epsilon_k^\text{Lyap} < \delta^\text{slack}$} {$\lambda_k^\text{Lyap} \leftarrow \alpha\lambda_k^\text{Lyap}$} {$\lambda_k^\text{Lyap} \leftarrow 0$}
    \State Solve~\eqref{eq:optf} w.r.t. $\{\valfunc_k, \mathbf \epsilon_k, \mathbf{\hat\lag}_k\}$
    \IfThenElse{$\epsilon_k^\text{Lyap} < \delta^\text{slack}$} {$\lambda_k^\text{Lyap} \leftarrow \alpha\lambda_k^\text{Lyap}$} {$\lambda_k^\text{Lyap} \leftarrow 0$}
    \State Solve~\eqref{eq:optf} w.r.t. $\{\fbctrl_k, \rho_k, \mathbf \epsilon_k, \mathbf{\hat \lag}_k\}$
    \IfThenElse{$\epsilon_k^\text{Lyap} < \delta^\text{slack}$} {$\lambda_k^\text{Lyap} \leftarrow \alpha\lambda_k^\text{Lyap}$} {$\lambda_k^\text{Lyap} \leftarrow 0$}   
    \IfThen{$\max |\text{coef}(\valfunc_k- \valfunc_k^\text{prev}, \rho_k- \rho_k^\text{prev}, \fbctrl_k- \fbctrl_k^\text{prev})| < \delta^\text{conv} \wedge \epsilon_k^\text{Lyap} < \delta^\text{slack}$}{exit loop} 
    \State $i \leftarrow i+1$.
    \EndWhile
    \State  \Return $(\valfunc_k, \rho_k, \fbctrl_k)$
  \end{algorithmic}
\end{algorithm} 

\begin{subequations}\label{eq:optf}
  \begin{align}
  \underset{\valfunc_k,\fbctrl_k, \rho_k, \mathbf{\lag}_k, \mathbf{\epsilon}_k}{\text{maximize}}\quad &\rho_k - \int_\roi\valfunc_k(\state)d\state - \mathbf{\lambda}_k \cdot \mathbf{\epsilon}_k\label{eq:optf:obj} \\
  \text{subject to}\quad & 
  \begin{aligned}
  &-\Big(\frac{\partial\valfunc(s,\state)}{\partial\state} \cdot \dyn(\state, \fbctrl_k(\state), \dstb) \\
  & \qquad + \frac{\valfunc_{k+1}(\state)-\valfunc_k(\state)}{s_{k+1} - s_k} - \frac{\rho_{k+1}-\rho_{k}}{s_{k+1} - s_k} - \epsilon_k^\text{Lyap} \Big) \\
  & \qquad\qquad\qquad + \lag_{\dset,k}^\text{Lyap}(\state,\dstb) \setfunc_\dset(\dstb)\\ 
  & \qquad\qquad\qquad + \lag_{\raset,k}^\text{Lyap}(\state,\dstb)(\valfunc_k(\state) - \rho_k)  \\ 
  & \qquad\qquad\qquad - \lag_{\targetset,k}^\text{Lyap}(\state,\dstb)\setfunc_\targetset(\state) \\
  & \qquad\qquad\qquad + \lag_{\roi,k}^\text{Lyap}(\state,\dstb) \setfunc_\roi(\state) \issos 
  \end{aligned} \label{eq:optf:lyap} \\
  & 
  \begin{aligned}
  \valfunc_k(\state) - \rho_k &- \lag_{\targetset,k}^\text{ca}(\state)\setfunc_\targetset(\state) \\
  &+\lag_{\avoidset,k}^\text{ca}(\state)\setfunc_\avoidset(\state) \\
  &+ \lag_{\roi,k}^\text{ca}(\state) \setfunc_\roi(\state) \issos 
  \end{aligned}\label{eq:optf:ca} \\
  & \begin{aligned}
  -\setfunc_\cset(\fbctrl_k(\state)) &+ \lag_{\raset,k}^\text{ctrl}(\state,\dstb)(\valfunc_k(\state) - \rho_k) \\
  &+ \lag_{\dset,k}^\text{ctrl}(\state,\dstb)\setfunc_\dset(\dstb) \\
  &+\lag_{\roi,k}^\text{ctrl}(\state,\dstb) \setfunc_\roi(\state)\issos
  \end{aligned} \label{eq:optf:ctrl_sat}\\
  & \begin{aligned}
  - (\valfunc_k(\state) - \rho_k) &+ \lag_{\targetset,k}^\text{it}(\state) (\valfunc_{k+1}(\state) - \rho_{k+1} + \epsilon_k^\text{it}) \\
  &+\lag_{\roi,k}^\text{it}(\state) \setfunc_\roi(\state) \issos
  \end{aligned} \label{eq:optf:it}\\  
  &\begin{aligned}
  &\valfunc_k(\state), \lag_{\dset,k}^\text{Lyap}, \lag_{\targetset,k}^\text{Lyap}, \lag_{\targetset,k}^\text{ca}, \lag_{\avoidset,k}^\text{ca}, \lag_{\raset,k}^\text{ctrl}, \lag_{\dset,k}^\text{ctrl}, \\
  &\qquad  \lag_{\targetset,k}^\text{it}, \lag_{\roi,k}^\text{Lyap}, \lag_{\roi,k}^\text{ca}, \lag_{\roi,k}^\text{ctrl}, \lag_{\roi,k}^\text{it} \text{ are SOS} 
  \end{aligned} \label{eq:optf:SOS} \\
  &\epsilon_k^\text{Lyap}, \epsilon_k^\text{it} \ge 0 \nonumber
  \end{align}
\end{subequations}

\noindent where $\mathbf\lambda_k = (\lambda_k^\text{Lyap},\lambda_k^\text{it})$ is a constant vector of weights, and
\begin{align*}
&\begin{aligned}
\mathbf L = \Big\{\lag_{\dset,k}^\text{Lyap}, \lag_{\raset,k}^\text{Lyap}, \lag_{\targetset,k}^\text{Lyap}, \lag_{\roi,k}^\text{Lyap}, \lag_{\targetset,k}^\text{ca}, \lag_{\avoidset,k}^\text{ca}, \lag_{\roi,k}^\text{ca}, \\
\lag_{\raset,k}^\text{ctrl}, \lag_{\dset,k}^\text{ctrl}, \lag_{\roi,k}^\text{ctrl}, \lag_{\targetset,k}^\text{it}, \lag_{\roi,k}^\text{it}\Big\}.
\end{aligned}\\
&\mathbf{\epsilon}_k = (\epsilon_k^\text{Lyap}, \epsilon_k^\text{it})
\end{align*}

\section{Numerical Example: the Reach-Avoid Game \label{sec:num}}
We now demonstrate our approach for computing reach-avoid sets by analyzing the reach-avoid game, which involves an \textit{attacker} with state $\astate$ trying to reach a target $\targetset^\text{A}$ while avoiding capture by the \textit{defender} with state $\dstate$. Let the joint state be denoted $\state = (\astate, \dstate)$. The joint dynamics are given in Eq. \eqref{eq:dyn}, where we use $\ctrl$ to model the control of the attacker, and $\dstb$ to model the control of the defender.

For convenience, we will use $\apos, \dpos$ to denote the positions of the attacker and the defender, respectively. In general $\apos,\dpos$ will be subsets of the player states $\astate, \dstate$, respectively.

In the context of reach-avoid problems, the target set is the set of joint states such that the attacker is at the target:

\begin{equation*}
\targetset = \{\state: \astate \in \targetset^\text{A}\}
\end{equation*}

The avoid set $\avoidset$ is the set of joint states in which the attacker is captured by the defender. In this paper we will assume that the attacker is captured by the defender when the positions of the two players are within some capture radius:

\begin{equation}
\avoidset = \{\state: \|\apos - \dpos\|_2 \le \capradius\}
\end{equation}


We now present reach-avoid set computations for two examples of reach-avoid games. The first example involves two single integrator players moving in 2D space; the joint state space dimension is 4D; we compare our computation results with those obtained from HJ reachability, which is the most general method that provides the optimal solution up to small numerical errors. The second example involves two kinematic car players moving in 2D; the joint state space dimension is 6D, and computation using HJ reachability is intractable.

\subsection{Two single integrator players}

Consider the following player dynamics:

\begin{align}
& \dastate_1 = \ctrl_1 & \ddstate_2 = \dstb_2 \\
& \dastate_2 = \ctrl_2 & \ddstate_2 = \dstb_2 \\
& |\ctrl_i| \le \bar\ctrl & |\dstb_i| \le \bar\dstb
\end{align}

Traditionally, the reach-avoid set for these player dynamics can be computed using HJ reachability \cite{Huang2015,Chen17}. 
Under special scenarios such as those in which players have the same maximum speed, analytic methods may be employed.
Like HJ reachability, our SOS-based approach is applicable in a general setting.
In this section, by comparing our results with those of HJ reachability, we demonstrate that our numerical results are conservative approximations of the optimal reach-avoid set, and therefore maintains reaching and avoidance guarantees.

For this example, we have chosen the maximum speeds to be $\bar\ctrl=2, \bar\dstb=1$, and the target set to be approximately a square of length $1$ centered at the origin, $\targetset^\text{A} = \{\astate: (\astate_1)^4 + (\astate_2)^4 \le 1\}$. 
The hyper-parameters of the optimization are $\delta^\text{slack} = 10^{-9}$, $\delta^\text{conv} = 10^{-3}$, $\lambda_k^\text{Lyap} = 5$, $\lambda_k^\text{it} = 20$, $\alpha = 1.5$. 
The maximum degree of $\valfunc_k$ was set to $4$, that of $\fbctrl_k$ set to $3$, maxIter set to $30$, the region of interest $\roi$ set to $[-3, 3]^4$, and the time discretizations set to $s_k = 0.25k - 2, k = 0, 1,2,\ldots,N=8$.

\begin{figure}
  \centering
  \includegraphics[width=0.6\columnwidth]{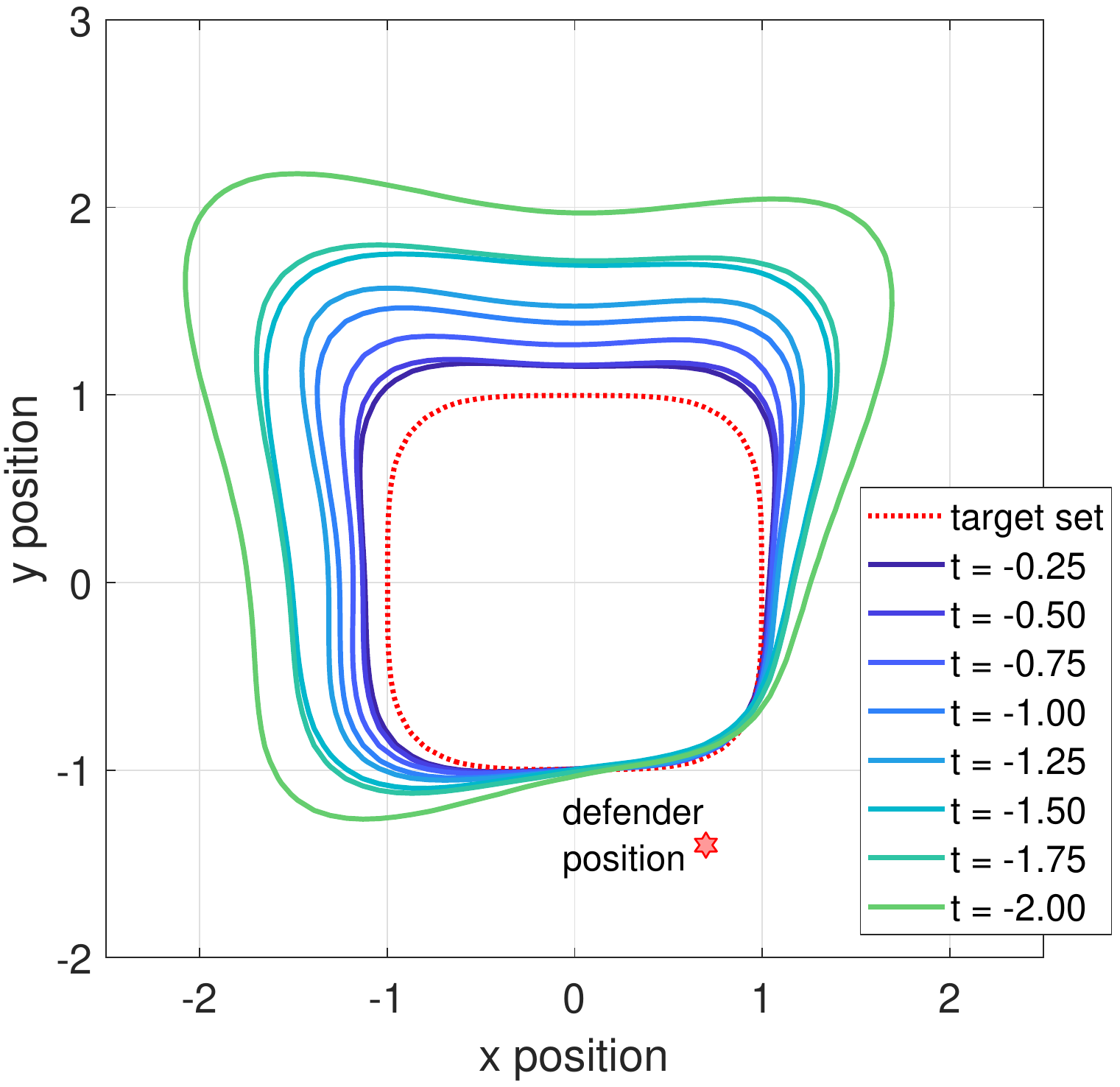}
  \caption{2D slice of the reach-avoid set $\raset(t)$ for the two single integrator players example at several values of $t$. The non-uniform growth of $\raset(t)$ over time is likely due to the suboptimal solutions of the SOS program.}
  \label{fig:int_set_over_time}
\end{figure}

To visualize the 4D reach-avoid set $\raset(t)$, we fix the defender position $\dstate$ and show 2D slices of $\raset(t)$ over several values of $t$ in Figure \ref{fig:int_set_over_time}. 
One can notice that the growth of $\raset(t)$ over time is not uniform as expected. 
This can be attributed to the solution of the SOS program being suboptimal, since the problem is non-convex.
However, it is important to note that given the constraints of the SOS program, any \textit{feasible} solution offers reaching and avoidance guarantees.

\begin{figure}
  \centering
  \includegraphics[width=0.8\columnwidth]{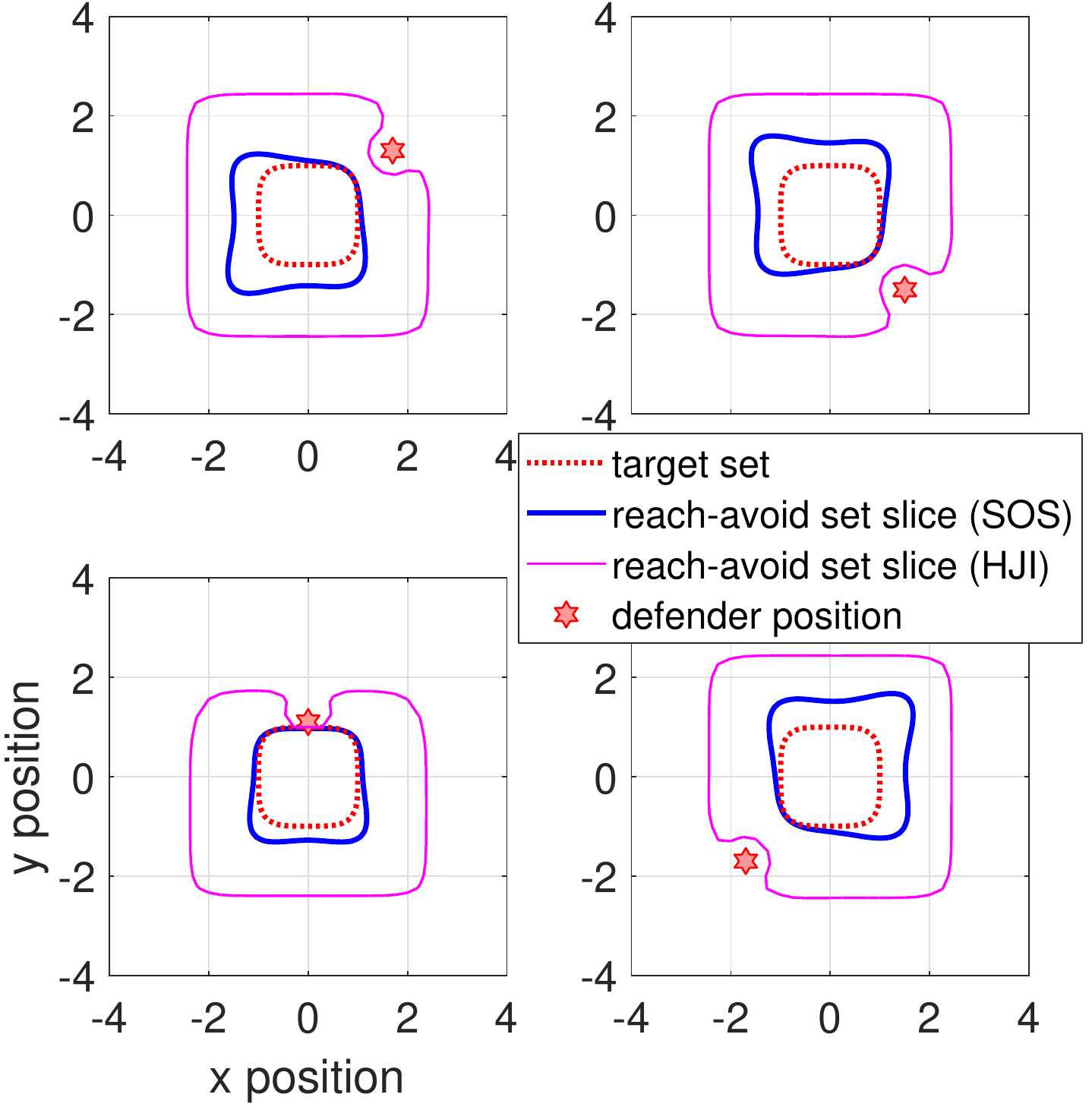}
  \caption{Comparison of reach-avoid set slices computed using our SOS-based method and using HJ reachability \cite{Chen17}. The reach-avoid set computed using our SOS-based method is a conservative approximation of the true reach-avoid set computed using HJ reachability.}
  \label{fig:int_set_slices}
\end{figure}

Figure \ref{fig:int_set_slices} shows computations of $\raset(t=1)$ sliced at various defender positions $\dstate$. 
The outer magenta boundary is the computation result from HJ reachability, and represents the true reach-avoid set up to small numerical errors. 
The solid blue boundary is the computation result from our SOS-based method. 
HJ reachability is better suited for this smaller 4D system, as the optimal solution can be obtained.
However, any state inside the set computed using SOS programming is inside the set computed using HJ reachability, which means our computation results, although conservative, maintain reaching and avoidance guarantees.

Computations for this example were done on a desktop computer with an Intel Core i7 2600K CPU and 16 GB of RAM. The SOS computations took approximately 17 minutes with the above parameters using the spotless toolbox \cite{spotless} and Mosek \cite{mosek2010mosek}, and the HJ computations took approximately 25 minutes on a grid with $45^4$ points using the level set toolbox \cite{Mitchell07c} and the helperOC library \cite{helperOC}. Computational time varies greatly with the maximum degree of polynomials chosen in the case of the SOS-based method, and with the number of grid points in the HJ method.

\subsection{Two kinematic car players}
In this section, we demonstrate our method on a system involving two kinematic cars. The joint dynamics are

\begin{align}
& \dastate_1 = v_A \cos \astate_3 & \ddstate_1 = v_D \cos \dstate_3\\
& \dastate_2 = v_A \sin \astate_3 & \ddstate_2 = v_D \sin \dstate_3\\
& \dastate_3 = \omega_A & \ddstate_3 = \omega_D \\
& 0 \le v_A \le \bar v_A, |\omega_A| \le \bar \omega_A & 0 \le v_D \le \bar v_D, |\omega_D| \le \bar \omega_D
\end{align}

Here we apply our SOS-based approach to derive and certify a controller for one of the cars that enables it to reach a goal in state space no matter what control action the other car might perform. 
This is indeed useful in the context of, for example, designing a safe lane-following controller that makes no assumption on the policy of other drivers, except for control bounds.
Even though our controller and reach-avoid set are once again quite conservative, they provide formal guarantees on a system that is typically too high-dimensional for treatment with state of the art approaches like HJ-based methods.

For this example, the maximum speed and turning rates of the attacker are $\bar v_A = 5$ and $\bar \omega_A = 3$. 
We assume that the other car, the defender, will not try to collide with us on purpose and is therefore limited to speeds of $\bar v_B = 3$ and $\bar \omega_B = 1$ for its velocity and turn rate.
Such an assumption can be made when one assumes, for example, that the other driver is simply trying to stay inside his lane.
The target set, representing our desire to have the attacker stay on the road without colliding with the other car, is a circle in the middle of the road ($\mathcal{T}^A = \{x^A: (x_1^A)^2 + (x_2^A)^2 \leq .25\}$) on the right side of the region of interest (a box from $-2$ to $0$ in $\astate_1$,$\dstate_1$ and from $-.5$ to $.5$ in $\astate_2$,$\dstate_2$).
We also bound $\astate_3$ and $\dstate_3$ to be inside the interval $[-\pi/4, \pi/4]$, which allows us to use a Chebyshev approximation with an accuracy of $1.5\times10^{-4}$ for $\sin\theta$ and $2\times10^{-3}$ for $\cos\theta$, namely:
\begin{align*}
& \sin(\theta) \approx 0.7264 (\theta/(\pi/4)) \nonumber\\
& \quad\quad\quad - 0.01942 (4 (\theta/(\pi/4))^3 - 3 (\theta/(\pi/4)))\\
& \cos(\theta) \approx 0.8516 - 0.1464 (2 (\theta/(\pi/4))^2 -1).
\end{align*}
The approximation makes our dynamics polynomial, and therefore amenable to SOS optimization. The hyper-parameters of the optimization are $\delta^\text{slack} = 10^{-5}$, $\delta^\text{conv} = 10^{-5}$, $\lambda_k^\text{Lyap} = 10$, $\lambda_k^\text{it} = 10$, $\alpha = 2$. The maximum degree of $\valfunc_k$ was set to $4$, that of $\fbctrl_k$ set to $3$. The maximum number of iterations (maxIter) was set to $10$, and the time discretizations set to $s_k = 0.1k - 1, k = 0, 1,2,\ldots,N=10$.

\begin{figure}
  \centering
  \includegraphics[width=0.49\columnwidth]{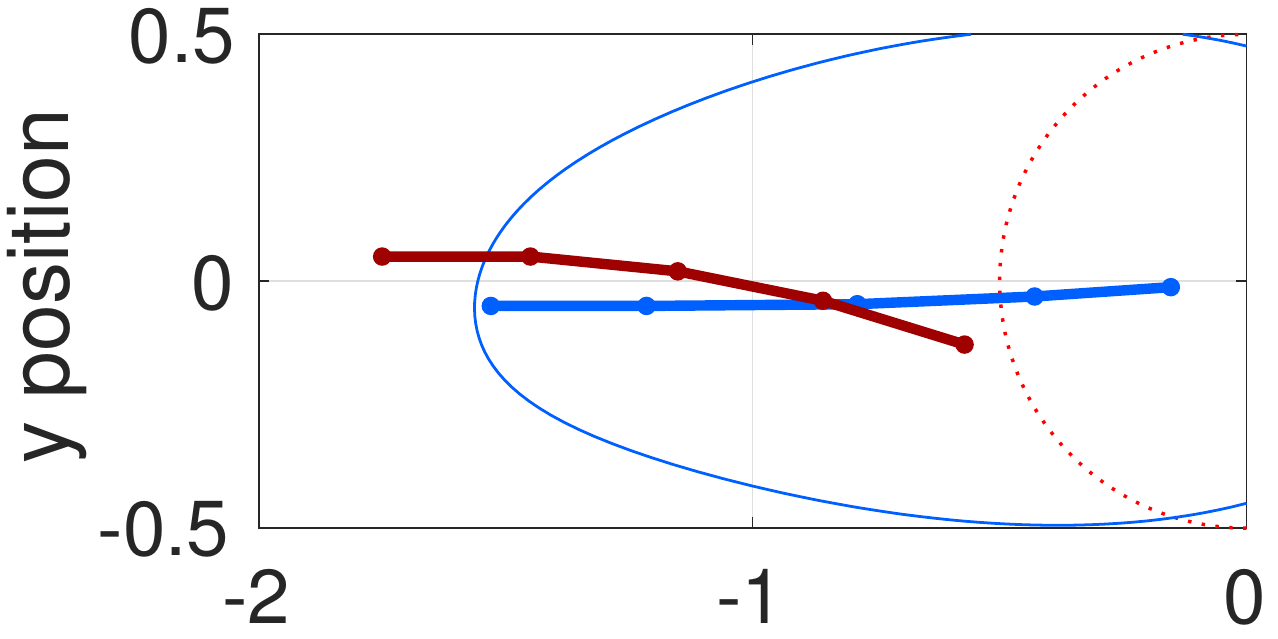}
  \includegraphics[width=0.49\columnwidth]{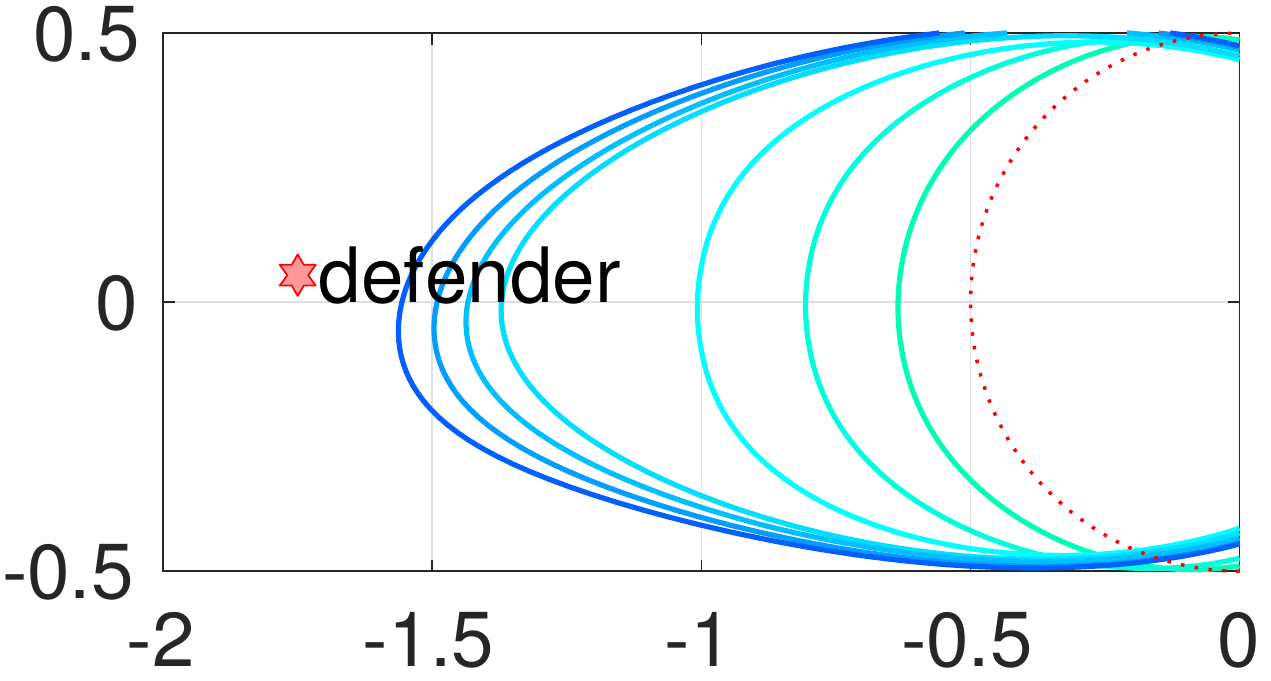}
  \includegraphics[width=0.49\columnwidth]{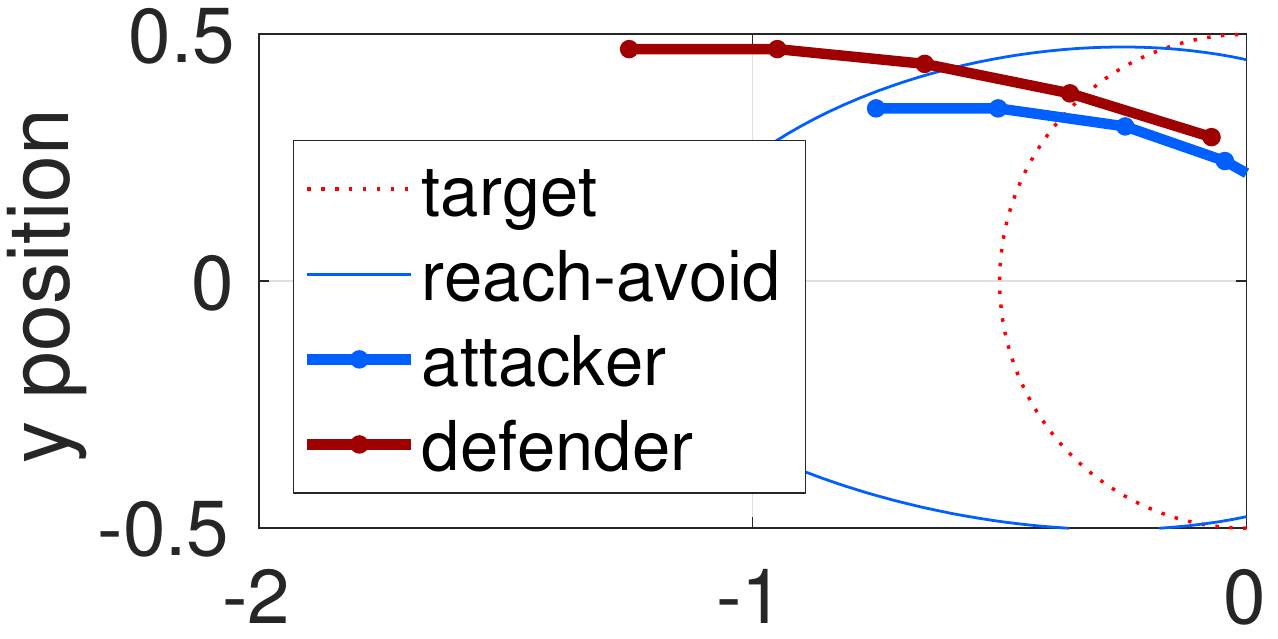}
  \includegraphics[width=0.49\columnwidth]{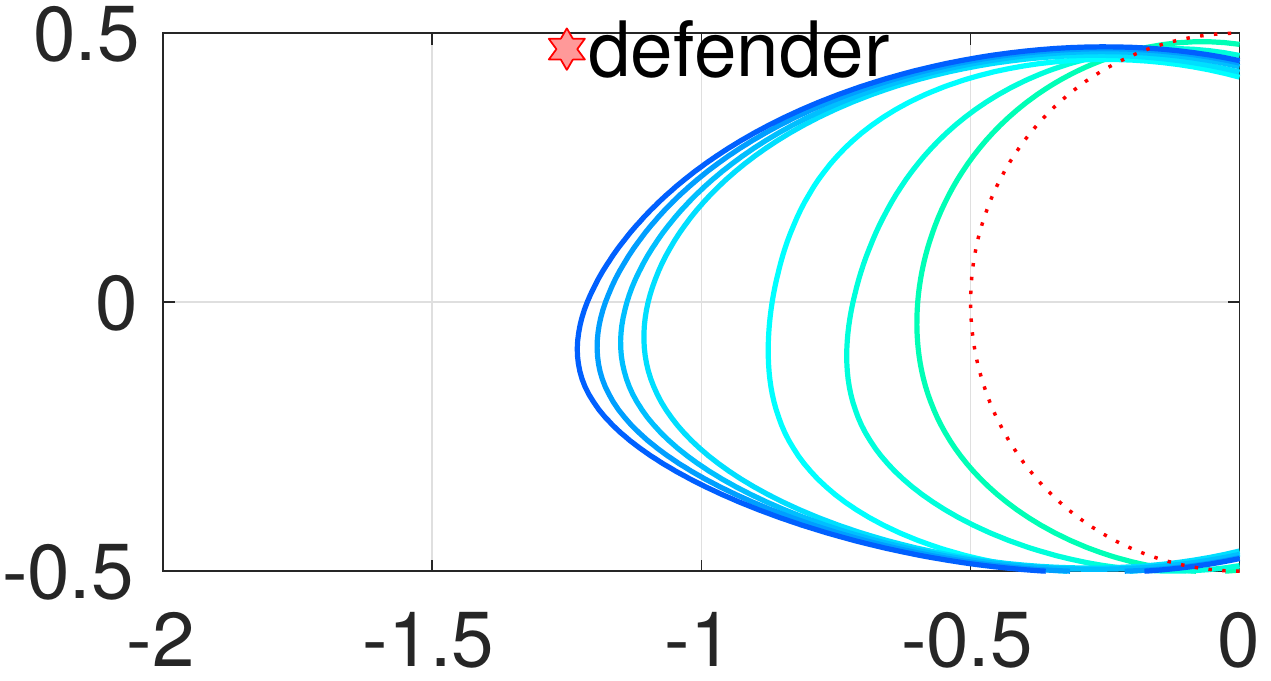}
  \includegraphics[width=0.49\columnwidth]{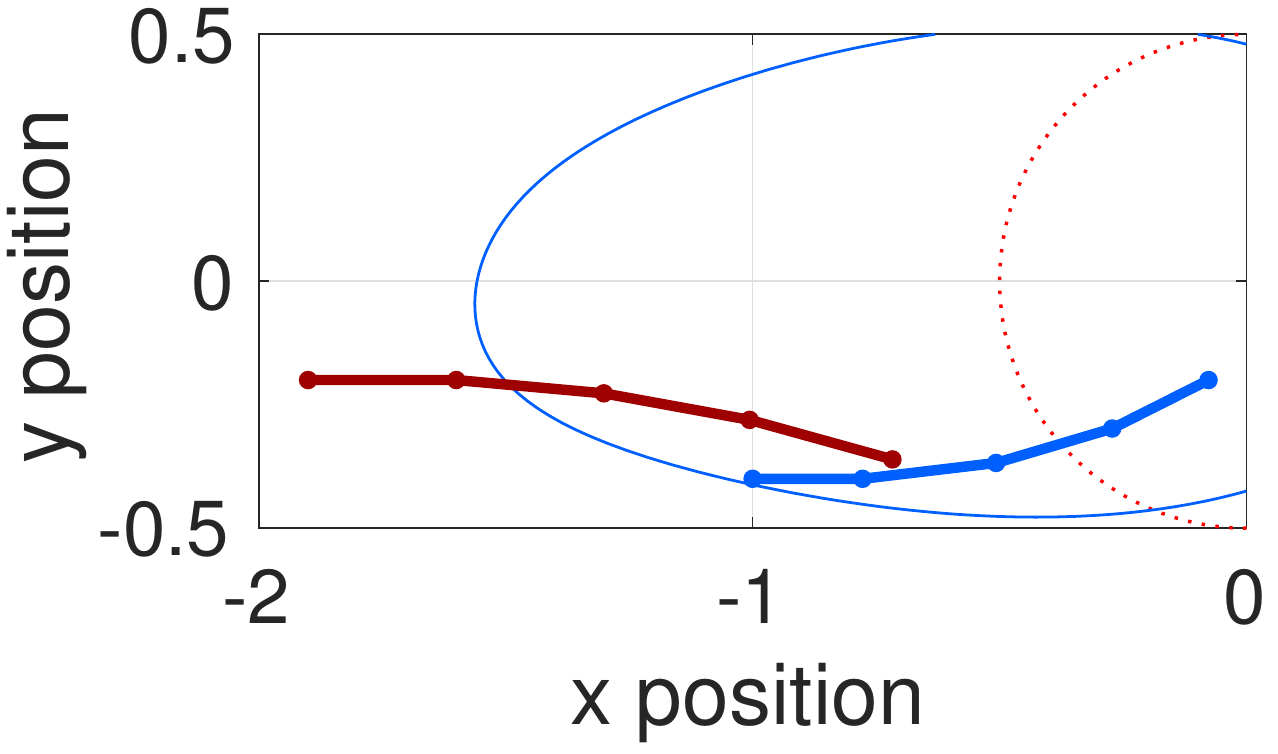}
  \includegraphics[width=0.49\columnwidth]{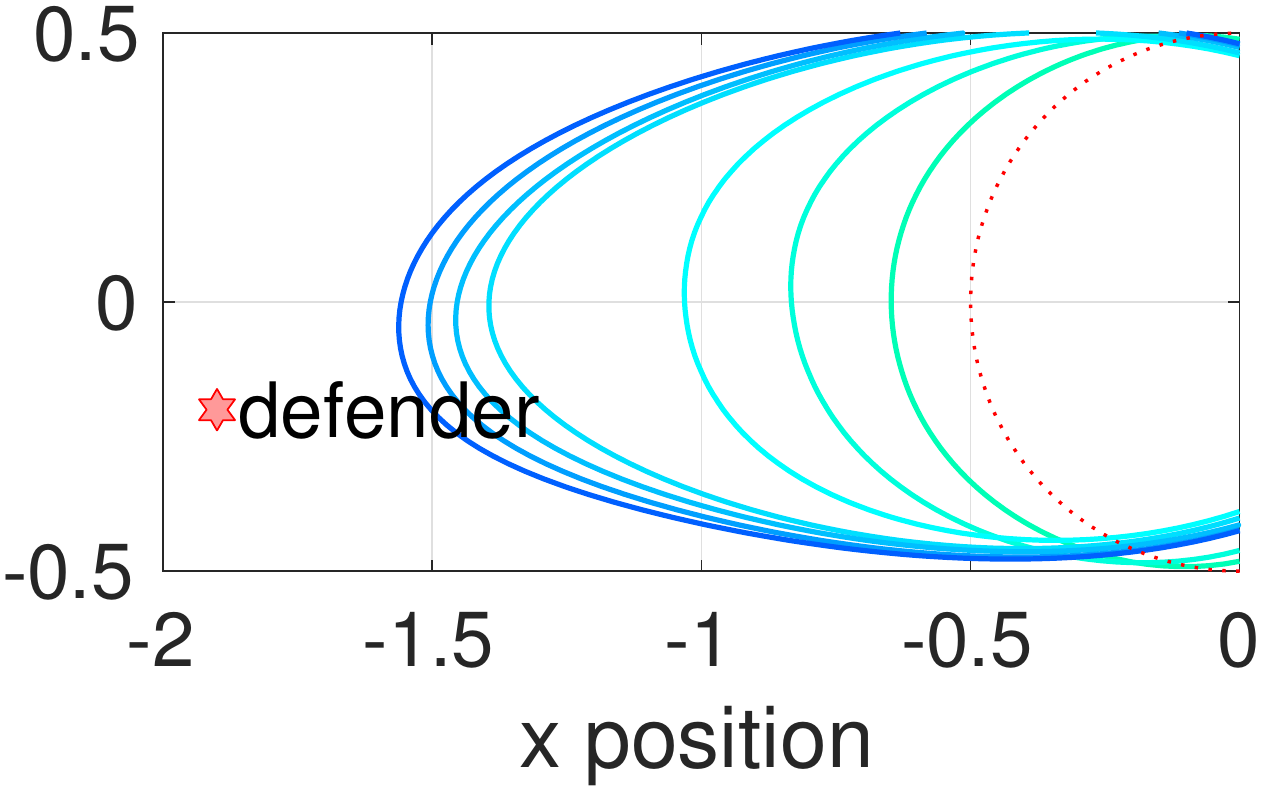}
  \caption{Left: the result of simulating the controller returned by our SOS-based approach for different initial states of the two cars. The attacker successfully navigates to the goal without colliding with the other car (the defender) each time it starts in the reach-avoid set $\raset(t)$. Right: 2D slice of the reach-avoid set $\raset(t)$ for the two kinematic cars example at several values of $t$.}
  \label{fig:cars_set_over_time}
\end{figure}

The right column of Figure \ref{fig:cars_set_over_time} shows slices of the reach-avoid slices for the states corresponding to $\astate_3 = 0$, $\dstate_3 = 0$, and the indicated defender position. 
The left column shows the result of applying the resulting controller in simulation. 
Even though the sets are conservative, none of the constraints are violated in simulation. 
Also note that the policy used by the defender is simply to maximize its velocity in some direction. 
Importantly though, the controller returned by our algorithm is robust to \textbf{any} policy the defender might use. A video of this example is available online\footnote{\url{https://www.youtube.com/watch?v=gUytdFHkjYY}}. 

Computation for this example were performed on a desktop computer with an AMD Ryzen 7 1800X Eight-Core Processor and 32 GB of RAM. The computations took approximately 4 hours using the Spotless toolbox \cite{spotless} and the Mosek solver \cite{mosek2010mosek} with little tuning on the convergence detection.
\section{Conclusions and Future Work}\label{sec:conclusion}
We presented a novel method for computing reach-avoid sets and synthesizing a feedback controller that guarantees reaching and avoidance when the system starts inside the set. 
Our method utilizes sum-of-squares (SOS) optimization to trade-off optimality of solution for computational complexity, allowing us to compute, for the first time to the best of our knowledge, 6D reach-avoid sets, although our solution is conservative.
Combining SOS optimization with dynamic programming, we also greatly reduce the computational complexity of solving the SOS program.

Future work includes investigating ways to reduce both the degree of conservatism in our solutions and the computation time by improving the optimization algorithm. 
In addition, since our method is applicable to polynomial system dynamics and general problem setups, there is much room for the exploration of potential applications of our work. 
Lastly, we would like to test our algorithm through hardware implementation. 

\section{Acknowledgements}\label{sec:acknowledgements}
The authors would like to thank Anirudha Majumdar for his guidance on sum-of-squares programming. This work was supported by the Office of Naval Research YIP program (Grant N00014-17-1-2433) and by the Toyota Research Institute (TRI). This article solely reflects the opinions and conclusions of its authors and not ONR, TRI or any other Toyota entity.







\bibliographystyle{IEEEtran}
\bibliography{landry.chen.hemley.pavone.iros18}

\end{document}